\documentclass[cits]{PoS} 
\usepackage{amsmath,amssymb,bm,booktabs,cite,color,cool}

\newcommand{\ups}{\upsilon}
\title{Semirelativistic Bound States: (Pseudo-) Spinless-Salpeter
Approaches Reassessed}\ShortTitle{Semirelativistic Bound States:
(Pseudo-) Spinless-Salpeter Approaches Reassessed}\author
{\speaker{Wolfgang Lucha}\\Institute for High Energy Physics,
Austrian Academy of Sciences, Nikolsdorfergasse 18,\\A-1050
Vienna, Austria\\E-mail: \email{Wolfgang.Lucha@oeaw.ac.at}}

\abstract{Relativistic quantum field theory offers, in form of the
homogeneous Bethe--Salpeter framework, a (Poincar\'e-covariant)
description of bound states in terms of their underlying theory's
fundamental degrees of freedom. In view of the intrinsic
complexity of this approach, simplifications have been sought and
abundantly found. The significance of these latter approximations
may be estimated by comparing their predictions with (easily
inferable) rigorous constraints on the bound-state spectra, such
as existence, number and location of discrete eigenstates. The
application of these techniques to selected proposed bound-state
equations is exemplified for a large class of generalizations of
the Hellmann potential frequently employed in several areas of
science such as physics and chemistry.}

\FullConference{European Physical Society Conference on High
Energy Physics -- EPS-HEP2019\\10--17 July, 2019\\Ghent, Belgium}

\begin{document}\section{Different Semirelativistic Bound-State
Treatments of Poincar\'e-Covariant Descent}Firmly grounded within
the framework of relativistic quantum field theories, the
homogeneous Bethe--Salpeter equation \cite{BS} forms a
Poincar\'e-covariant, albeit not always easy-to-handle, approach
to bound states. Driven by the desire to obtain analytic, thus
easier to control, yet still to some extent (semi-) relativistic
bound-state treatments, more or less severe simplifications of the
Bethe--Salpeter formalism have been proposed: Ignoring entirely
all dependence on timelike variables generates the rather broad
class of (merely) \emph{instantaneous\/} Bethe--Salpeter equations
\cite{WLe}. Assuming, furthermore, also free propagation of the
bound-state constituents leads to the \emph{Salpeter equation\/}
\cite{SE}. Skipping all negative-energy contributions and all
reference to any spin degrees of freedom eventually yields the
\emph{spinless Salpeter equation\/}, the eigenvalue equation of a
Hamiltonian $H$ generically consisting of the bound-state
constituents' relativistic kinetic energy and a potential $V$
encoding all their interactions. For the case of bound states of
two particles of equal masses $m$, each such Hamiltonian $H$ thus
reads\begin{equation}H\equiv2\,\sqrt{\bm{p}^2+m^2}+V(\bm{x})\
.\label{H}\end{equation}

The nonlocality of this operator $H$ renders hard to find
\emph{exact analytic\/} solutions to its \mbox{eigenvalue}
problem. In view of this, (rather bizarre) approximations to the
spinless Salpeter equation have~been proposed, by manipulations
such as expanding kinetic energies to one order beyond the
Schr\"odinger limit, arriving at operators not bounded from below,
and inserting the Schr\"odinger limit into the thus fabricated
\emph{pseudo-spinless-Salpeter\/} equations, entirely ignoring the
operator nature of the problem. Unsurprisingly, most of these
pseudo-spinless-Salpeter attempts do not withstand rigorous
scrutiny.

Definitely more reliable strategies rely on the derivation of
rigorous statements on the spectrum of the operator $H$, such as
establishing its boundedness from below (Sect.~\ref{sB}) or
providing bounds on number (Sect.~\ref{sN}) and location
(Sect.~\ref{sV}) of its discrete eigenvalues or validating
approximate findings by the proper \emph{relativistic virial
theorem\/} \cite{WL:RVT,WL:RVTs}. The latter bulk of tools has
been applied to relativistic problems defined by, for instance,
interaction potentials $V(\bm{x})=V(r)$, with $r\equiv|\bm{x}|$,
of \mbox{Woods--Saxon} \cite{WL:WS,WL:Q@W}, Hulth\'en
\cite{WL:Q@W,WL:H}, Yukawa \cite{WL:Y}, kink-like \cite{WL:K}, and
generalized-Hellmann \cite{WL:Hm}~form. We highlight the
capability of this programme for the illustrative set of
generalized Hellmann potentials (Sect.~\ref{sH}).

\begin{table}[hbt]\centering\caption{Classification of all
generalized Hellmann potentials $V_{\rm H}(r)$ with respect to the
size of the coupling $\ups$ of their Yukawa contributions relative
to the nonvanishing coupling $\kappa\gneqq0$ of their Coulomb
contribution \cite{WL:Hm}.}\label{GHP}\vspace{1ex}\begin{tabular}
{lllcc}\toprule Boundedness$\quad$&Characteristic$\quad$&Behaviour
near&\multicolumn{1}{l}{Sign of sum}&\multicolumn{1}{l}{Relation
between}\\ from below&of potential&the origin $r=0\quad$&
\multicolumn{1}{l}{of couplings}$\quad$&\multicolumn{1}{l}
{couplings $\ups$ and $\kappa$}\\\midrule unbounded&``singular''&
$V_{\rm H}(r)\xrightarrow[r\to0]{}-\infty$&$\kappa+\ups>0$&
$\ups>\kappa$\\&&&&$\ups=\kappa$\\&&&&$0<\ups<\kappa$\\&&&&
$\ups=0$\\&&&&$-\kappa<\ups<0$\\\midrule bounded&finite at origin&
$V_{\rm H}(r)\xrightarrow[r\to0]{}\ups\,b$&$\kappa+\ups=0$&
$\ups=-\kappa$\\&repulsive core&$V_{\rm H}(r)\xrightarrow[r\to0]{}
+\infty$&$\kappa+\ups<0$&$\ups<-\kappa$\\\bottomrule\end{tabular}
\end{table}

\section{Set of Generalized Hellmann-type Potentials:
Classification by Overall Behaviour}\label{sH}
\begin{figure}[hb]\begin{center}\begin{tabular}{ccc}
\includegraphics[scale=1.32401]{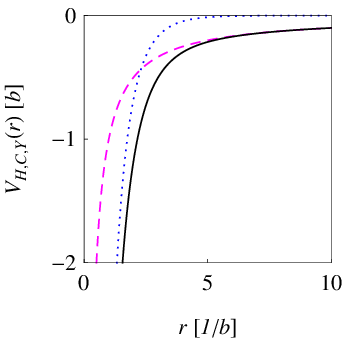}&
\includegraphics[scale=1.32401]{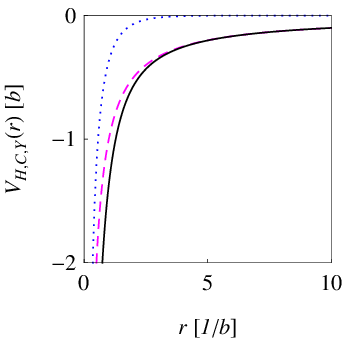}&
\includegraphics[scale=1.32401]{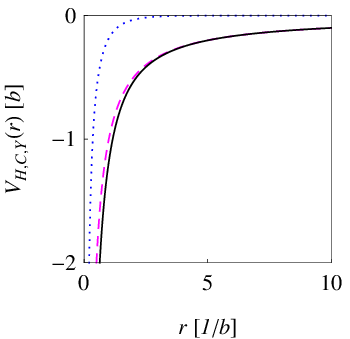}\\[1ex]
(a)&\textcolor{red}{(b)}&(c)\\[2ex]
\includegraphics[scale=1.32401]{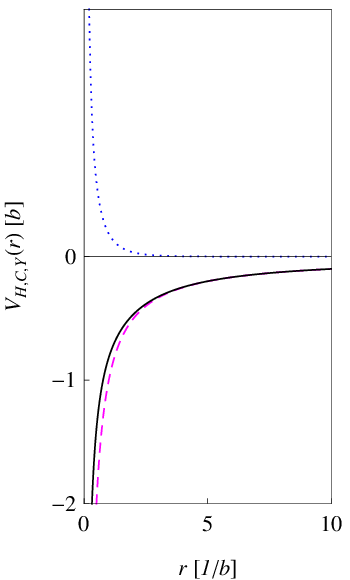}&
\includegraphics[scale=1.32401]{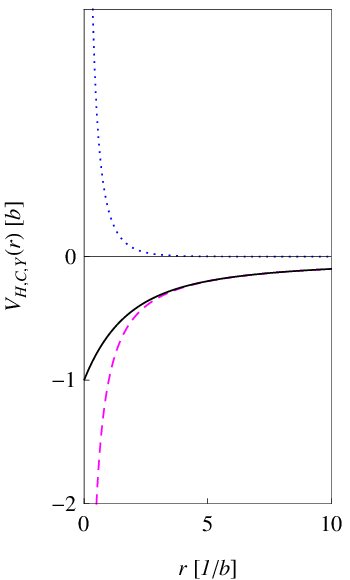}&
\includegraphics[scale=1.32401]{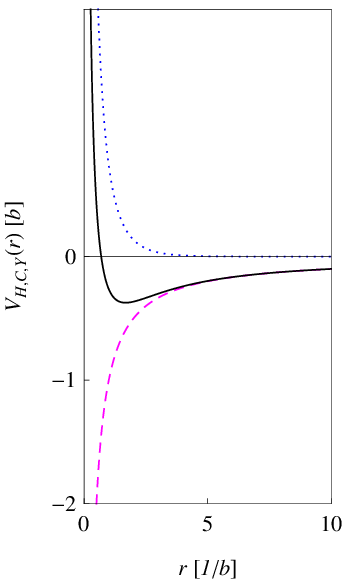}\\[1ex]
(d)&\textcolor{red}{(e)}&\textcolor{red}{(f)}\end{tabular}\caption
{Samples of the six \emph{non-Coulombic\/} types of (solid black)
\emph{generalized\/} Hellmann potentials $V_{\rm H}(r)$~of Table
\protect\ref{GHP} \cite{WL:Hm}, consisting of a (dashed magenta)
Coulomb part $V_{\rm C}(r)$, with coupling $\kappa=1$, plus a
(dotted blue) Yukawa part $V_{\rm Y}(r)$, with coupling $\ups=10$
(a), $\ups=1$ (b), $\ups=0.5$ (c), $\ups=-0.5$ (d), $\ups=-1$ (e),
or $\ups=-2$~(f).}\label{RGHP}\end{center}\end{figure}

Broadening
Hellmann's idea \cite{H,HK} of the potential experienced by
valence electrons in~atoms to involve attractive Coulomb and
repulsive Yukawa portions, the set of all superpositions
(Fig.~\ref{RGHP})~of a Coulomb contribution, with nonnegative
coupling $\kappa$, and a Yukawa-like contribution with positive
range parameter $b$ and coupling $\ups$ of either sign defines the
class of generalized Hellmann~\mbox{potentials}
\begin{equation}V_{\rm H}(r)\equiv V_{\rm C}(r)+V_{\rm Y}(r)
=-\frac{\kappa}{r}-\ups\,\frac{\exp(-b\,r)}{r}\ ,\qquad\kappa\ge0\
,\qquad\ups\gtreqqless0\ ,\qquad b>0\ .\label{V}\end{equation}Both
individual components, tantamount to its \emph{parametric
limits\/} $\ups\to0$, the Coulomb potential, and $\kappa\to0$, the
Yukawa potential, have been discussed thoroughly in
Refs.~\cite{IWH,IWHa,MR,RRSMS,WL:C} and \cite{WL:Y}, respectively.

The types of arising spectra allow to subdivide this set into the
seven categories of Table~\ref{GHP} \cite{WL:Hm}.

\section{Eigenvalue Spectra and Coupling Parameters:
Straightforward Rigorous Bounds}\label{sB}The first goal in one's
quest for bound states must be to identify constraints on the
potential that ensure the semiboundedness of the spectrum
$\sigma_H$ of the resulting spinless relativistic Hamiltonian $H$.
For large distances $r$, every potential of generalized Hellmann
shape approaches its Coulombic~part:$$V_{\rm H}(r)\xrightarrow
[r\to\infty]{}V_{\rm C}(r)\ .$$This observation, in turn, entails
that all discrete eigenvalues $E_k$ of $H$, related to the bound
states, are bounded from above by $E_k\le0,$ $k=0,1,2,\dots$, in
other words, all of these are definitely nonpositive.
\begin{description}\item[``Nonsingular'' Hellmann potentials]are
bounded from below [since they all satisfy $V_{\rm H}(r)>-\infty$]
and thus, due to the nonnegativity $\sqrt{\bm{p}^2+m^2}\ge0$ of
any kinetic term, also their Hamiltonians:
\begin{alignat}{3}H\ge V_{\rm H}(r)&\ge\min_{0\le r<\infty}V_{\rm
H}(r)>-\infty&\qquad\mbox{for}\qquad&\ups<-\kappa\ ,&\label{Sf}
\\[.3ex]H\ge V_{\rm H}(r)&\ge V_{\rm H}(0)=\ups\,b&\qquad\mbox{for}
\qquad&\ups=-\kappa\ .&\label{Se}\end{alignat}\item[``Singular''
Hellmann potentials]develop, because of the relation
$\ups>-\kappa$ among their couplings, negative singularities at
$r=0$ bounded from below by an associated Coulomb-like behaviour:
$$V_{\rm H}(r)\ge-\frac{\alpha}{r}\qquad\mbox{with}\qquad
\left\{\begin{array}{ll}\alpha=\kappa+\ups&\qquad\mbox{for}\qquad
\ups>0\qquad\Longleftrightarrow\qquad V_{\rm Y}(r)<0\ ,\\[.5ex]
\alpha=\kappa&\qquad\mbox{for}\qquad\ups\le0\qquad
\Longleftrightarrow\qquad V_{\rm Y}(r)\ge0\ .\end{array}\right.$$
The spectra of the thereby defined relativistic Coulomb problems,
$\sigma_H$, satisfy \cite{IWH,MR}, in~turn,
\begin{equation}\sigma_H\ge2\,m\times\left\{\begin{array}{ll}
\displaystyle\sqrt{1-\left(\frac{\pi\,\alpha}{4}\right)^{\!2}}&
\displaystyle\qquad\mbox{if and only if}\qquad\alpha\le\frac{4}
{\pi}=1.273239\dots\ ,\\[2.2ex]\displaystyle\sqrt{\frac
{1+\sqrt{1-\alpha^2}}{2}}&\qquad\mbox{for}\qquad\alpha\le1\ ,
\end{array}\right.\label{SC}\end{equation}proving their Hellmann
counterparts to be, too, bounded from below. Exploiting the
trial~state$$\phi(\bm{x})\propto\exp(-\mu\,r)\qquad
\Longleftrightarrow\qquad\widetilde\phi(\bm{p})\propto\frac{1}
{(\bm{p}^2+\mu^2)^2}$$with $\mu>0$ in the expectation values of
both kinetic term $\sqrt{\bm{p}^2+m^2}$ \cite{WL:H,WL:Y} and
potential~$V_{\rm H}(r)$,\begin{align*}\left\langle\!\!
\sqrt{\bm{p}^2+m^2}\right\rangle&=\frac{2}{3\,\pi\,(m^2-\mu^2)^{5/2}}
\left[\mu\,\sqrt{m^2-\mu^2}\,(3\,m^4-4\,m^2\,\mu^2+4\,\mu^4)\right.\\
&+\left.3\,m^4\,(m^2-2\,\mu^2)\,\ArcSec{\mbox{\Large$\frac{m}{\mu}$}}
\right],\\[1ex]\langle V_{\rm H}(r)\rangle&=-\kappa\,\mu-
\frac{4\,\ups\,\mu^3}{(b+2\,\mu)^2}\ ,\end{align*}it is trivial to
establish that the couplings entering any Hellmann potential have
to be bounded:\begin{align*}\langle H\rangle&=\left(\frac{16}
{3\,\pi}-\kappa-\ups\right)\mu+O(1)\xrightarrow[\mu\to\infty]{}-
\infty\qquad\mbox{for}\qquad\kappa+\ups>\frac{16}{3\,\pi}\\[1ex]&
\Longrightarrow\qquad\left\{\begin{array}{rllr}\kappa+\ups\stackrel
{!}{\le}&\displaystyle\!\!\frac{16}{3\,\pi}=1.69765\dots&\qquad
\mbox{for}&\qquad\ups>0\ ,\\[2.2ex]\kappa\stackrel{!}{\le}&
\displaystyle\!\!\frac{16}{3\,\pi}+|\ups|&\qquad\mbox{for}&\qquad
-\kappa<\ups<0\ .\end{array}\right.\end{align*}\end{description}

\section{Playing the Mini--Max Game: Variational Upper Bounds on
Discrete Eigenvalues}\label{sV}As soon as the boundedness from
below of a reasonably defined Hamiltonian operator has been
established, pinning down its bound states does make sense. On the
basis of a characterization~of its discrete spectrum by the famous
minimum--maximum theorem, rigorous upper bounds on as~well as
improvable approximate estimates of the localization of
bound-state energy levels can be derived by a straightforward
application of variational techniques. In terms of generalized
Laguerre orthogonal polynomials $L_k^{(\gamma)}(x)$ of parameter
$\gamma$ and spherical harmonics ${\cal Y}_{\ell m}(\Omega)$ of
orbital angular momentum $\ell$ and projection $m$ depending on
the solid angle $\Omega$, a rather convenient trial-space basis
reads \cite{WL:T1,WL:T2,WL:Tr}
\begin{align}\psi_{k,\ell m}(\bm{x})\propto r^{\ell+\beta-1}
\exp(-\mu\,r)\,L_k^{(2\,\ell+2\,\beta)}(2\,\mu\,r)\,{\cal Y}_{\ell
m}(\Omega_{\bm{x}})\ ,\qquad k\in{\mathbb N}_0\ ,&\nonumber\\[1ex]
\;L_k^{(\gamma)}(x)\equiv\sum_{t=0}^k\,\binom{k+\gamma}{k-t}
\frac{(-x)^t}{t!}\ ,\qquad\mu>0\ ,\qquad\beta>-\frac{1}{2}\
.&\label{L}\end{align}Table~\ref{F} shows bounds on energy levels
for three of the six Yukawa-part-affected categories
in~Fig.~\ref{RGHP}.

\begin{table}[hbt]\centering\caption{Bounds on the binding energies
of low-lying eigenstates (identified by radial, $n_r$, and orbital
angular momentum, $\ell$, quantum number) of the Hamiltonian
(\protect\ref{H}) with generalized Hellmann potential
(\protect\ref{V}) for three illustrative choices of the couplings
$\kappa$ and $\ups$, and one common Yukawa range $b=m$. The upper
bounds result from our trial-space basis (\protect\ref{L}) for
parameters $\mu=m$ and $\beta=1$, the lower from
Eqs.~(\protect\ref{Sf}) through (\protect\ref{SC})~\cite{WL:Hm}.}
\label{F}\vspace{1.18ex}\begin{tabular}{cc|lll}
\toprule&&\multicolumn{3}{c}{Upper bound on $B_k\equiv
E_k-2\,m\;[m]$ ($k\in{\mathbb N}_0$)}\\\cline{3-5}\\[-2ex]
\multicolumn{2}{c|}{Bound state}&
\multicolumn{1}{c}{$\quad\kappa=\ups=\frac{1}{2}\quad$}&
\multicolumn{1}{c}{$\kappa=1$, $\ups=-1$}&
\multicolumn{1}{c}{$\quad\kappa=1$,
$\ups=-2\quad$}\\\cline{1-2}\\[-2ex]
$n_r$&$\ell$&\multicolumn{1}{c}{[Fig.~\protect\ref{RGHP}(b)]\quad}&
\multicolumn{1}{c}{[Fig.~\protect\ref{RGHP}(e)]}&
\multicolumn{1}{c}{[Fig.~\protect\ref{RGHP}(f)]}\\\midrule
0&0&$\quad-0.11673$&$\quad-0.17951$&$\qquad-0.14410$\\
0&1&$\quad-0.01579$&$\quad-0.06294$&$\qquad-0.06157$\\
0&2&$\quad-0.00616$&$\quad-0.02813$&$\qquad-0.02812$\\
0&3&$\qquad$~~---&$\quad-0.01553$&$\qquad-0.01553$\\[.5ex]
1&0&$\quad-0.02107$&$\quad-0.05464$&$\qquad-0.04786$\\
1&1&$\quad-0.00509$&$\quad-0.02810$&$\qquad-0.02762$\\
1&2&$\qquad$~~---&$\quad-0.01482$&$\qquad-0.01481$\\
1&3&$\qquad$~~---&$\quad-0.00624$&$\qquad-0.00624$\\[.5ex]
2&0&$\quad-0.00688$&$\quad-0.02566$&$\qquad-0.02338$\\
2&1&$\qquad$~~---&$\quad-0.01391$&$\qquad-0.01356$\\
2&2&$\qquad$~~---&$\quad-0.00122$&$\qquad-0.00120$\\[.5ex]
3&0&$\qquad$~~---&$\quad-0.01104$&$\qquad-0.00840$\\
\midrule\multicolumn{2}{l|}{Lower bound on $B_k\;[m]$}&
$\quad-0.58578\dots$&$\quad-1$&$\qquad-0.37336\dots$\\\bottomrule
\end{tabular}\end{table}

The quality of the corresponding approximate variational
eigenstates may be estimated by their fulfilment of the adequate,
i.e., \emph{relativistic\/} virial theorem \cite{WL:RVT,WL:RVTs}
pertaining to any exact eigenstate $|\chi\rangle$ of an operator
$H$, defined, of course, by $H\,|\chi\rangle=E\,|\chi\rangle$. For
our Hamiltonian (\ref{H}), the theorem~reads$$\left\langle
\chi\left|\frac{2\,\bm{p}^2}{\sqrt{\bm{p}^2+m^2}}\right|\chi
\right\rangle=\left\langle\chi\left|\,\bm{x}\cdot\frac{\partial\,V}
{\partial\bm{x}}(\bm{x})\right|\chi\right\rangle.$$

\section{Discrete Spinless-Salpeter Energy Levels: Constraining
Their Maximum Number}\label{sN}A central issue in any
spinless-Salpeter business is the actual \emph{number\/} of bound
states supported by a potential: how many bound states can one
expect to find? In particular, one would like to know, at least:
is their number finite or infinite? Unfortunately, in that context
exact results are not~abound.

A strict bound \cite{ICD} on the number of spinless-Salpeter bound
states exists for every nonpositive potential $V(\bm{x})$,
$V(\bm{x})\leq0$, satisfying the constraint (merely guaranteeing
the finiteness of this bound)$$V(\bm{x})\in L^{3/2}({\mathbb
R}^3)\cap L^3({\mathbb R}^3)\ .$$For a rather simple reason,
however, none of the generalized Hellmann potentials may belong to
this set: for large $r$, due to the rapid decay of its Yukawa part
any such potential approaches a Coulombic behaviour; the
corresponding number of discrete energy eigenvalues thus will grow
beyond bounds. Even if so, the spectral comparison theorem
recalled in Refs.~\cite{WL:C,WL:T1,WL:T2,WL:Tr,WL:RCP} offers
upper bounds on the individual spinless-Salpeter energy levels,
given by their nonrelativistic (Schr\"odinger) counterparts.

\end{document}